# Evaluating LLMs in the Context of a Functional Programming Course: A Comprehensive Study

Yihan Zhang[a] , Brigitte Pientka[a] , and Xujie Si[b]

a  McGill University, Montreal, Canada
b  University of Toronto, Toronto, Canada

**Abstract**    Large-Language Models (LLMs) are changing the way learners acquire knowledge outside the classroom setting. Previous studies have shown that LLMs seem effective in generating to short and simple questions in introductory CS courses using high-resource programming languages such as Java or Python.

In this paper, we evaluate the effectiveness of LLMs in the context of a low-resource programming language — OCaml, in an *educational* setting. In particular, we built three benchmarks to comprehensively evaluate 9 state-of-the-art LLMs: 1) λCODEGEN (a benchmark containing natural-language homework programming problems); 2) λREPAIR (a benchmark containing programs with syntax, type, and logical errors drawn from actual student submissions); 3) λEXPLAIN (a benchmark containing natural language questions regarding theoretical programming concepts). We grade each LLMs responses with respect to correctness using the OCaml compiler and an autograder. And our evaluation goes beyond common evaluation methodology by using manual grading to assess the quality of the responses.

Our study shows that the top three LLMs are effective on all tasks within a typical functional programming course, although they solve much fewer homework problems in the low-resource setting compared to their success on introductory programming problems in Python and Java. The strength of LLMs lies in correcting syntax and type errors as well as generating answers to basic conceptual questions. While LLMs may not yet match dedicated language-specific tools in some areas, their convenience as a one-stop tool for multiple programming languages can outweigh the benefits of more specialized systems.

We hope our benchmarks can serve multiple purposes: to assess the evolving capabilities of LLMs, to help instructors raise awareness among students about the limitations of LLM-generated solutions, and to inform programming language researchers about opportunities to integrate domain-specific reasoning into LLMs and develop more powerful code synthesis and repair tools for low-resource languages.

## The Art, Science, and Engineering of Programming

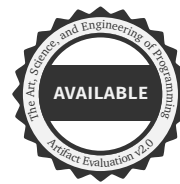 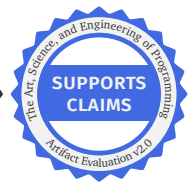

## 1 Introduction

Large Language Models (LLMs) are easily accessible to programmers and students. Students use LLMs to write code, fix code [30, 41, 57] as well as explain programs and coursework concept [20, 25, 28, 34, 42]. Whether we like it or not, LLMs are changing the way students acquire knowledge and seek explanations.

Yet, LLMs are a two-edged sword for a student. While LLMs seem effective in short and simple answers, they can also produce faulty ones that students think are correct. Kabir, Udo-Imeh, Kou, and Zhang [24] found that over 50 % of answers generated by ChatGPT to programming questions on Stack Overflow were incorrect and over 77 % of the generated answers were excessively verbose and not concise. However, over 35 % of the time, students often still preferred ChatGPT answers due to their comprehensive and well-articulated language style, and they overlooked the misinformation. This over-reliance on LLMs may pose challenges to students and instructors especially in more advanced courses. Not only will students find fewer resources online on some of these topics, but LLMs may also not be effectively trained due to the scarcity of data.

In this paper, we aim to understand the effectiveness of LLMs in the context of a 2nd year functional programming course that uses a low-resourced programming language – OCaml. This course covers basic to advanced functional programming techniques and introduces students to theoretical concepts in programming language design. This is in contrast to existing studies that have mostly focused on the use of mainstream languages in introductory computer science courses [12].

We focus on the following five research questions:

**RQ 1** Are LLMs solving functional programming assignments correctly, and concisely?

**RQ 2** Are LLMs repairing syntax, type and logical errors in functional programs correctly and concisely?

**RQ 3** Are LLMs explaining theoretical concepts regarding programming language design correctly and concisely?

**RQ 4** Do the LLMs' abilities to generate correct and concise answers vary across different types of questions – from generating and repairing code to explaining theoretical concepts?

**RQ 5** How does the difficulty level of a question impact the LLMs response?

To answer these questions we built three different benchmarks: 1) λCODEGEN (a benchmark containing natural-language homework programming problems), 2) λREPAIR (a benchmark containing programs with syntax, type, and logical errors drawn from actual student submissions) and 3) λEXPLAIN (a benchmark containing natural language questions regarding theoretical programming concepts). Our benchmark is much smaller compared to the commonly-used datasets such as HumanEval [9] and MBPP [3] because we also quantitatively and manually evaluate the responses generated by the LLM. Despite its size, it remains representative, as it captures key paradigms in functional programming.

The first benchmark λCODEGEN contains 10 *multi-task* programming assignments where each of them consists of several programming tasks given in natural language description and type specifications. They range from covering basic concepts such as pattern matching, data-types, basic higher-order functions such as map to more advanced programming techniques such as continuations. They also cover advanced concepts in programming languages such as programs that rely on the formal definition of bound variable, substitution, static and operational semantics for a toy language. As a result, our benchmark targets specific functional programming tasks rather than general programs in high-resources languages like Python, Java, or C.

The second benchmark λREPAIR contains 150 OCaml programs drawn from actual student errors in a 2nd year functional programming course at McGill University. It covers three different kinds of errors: syntax error (λREPAIRSYNTAX (50)), type error (λREPAIRTYPE (50)), and logical error (λREPAIRPROG (50)). This is in contrast to benchmarks such as QuickBugs [56] that use synthesized data.

Our last benchmark λEXPLAIN contains 50 questions extracted from the corpus of actual prep and exam questions focusing on (functional) programming concepts. These questions focus on concepts such as overwriting, overshadowing, call-by-value vs call-by-name evaluation vs evaluation in the presence of state, etc.

We use these benchmarks to assess how well LLMs perform on different tasks – from generating code (**RQ1**), repairing buggy code (**RQ2**), and providing explanations (**RQ3**). Furthermore, we compare the capabilities of LLMs across different types and difficulty levels of questions (**RQ4** and **RQ5**).

To assess the correctness of generated and repaired code, we use the OCaml compiler and an autograder. In addition, we conduct a manual assessment to evaluate not only the *correctness* but also the *quality* of the model output.

In our study, we consider 9 state-of-the-art LLMs, including widely-used models like GPT-4o. Recent surveys and studies show that ChatGPT/GPT-4o remains the most commonly used AI tool for students[1] ([17]). We observe that GPT-4o consistently performs in the top tier.

On λCODEGEN, top models like GPT-4o and o3-mini generate code at a Mastery level 74 % of the time, which are much higher than the results in earlier studies. The widely-used GPT-4o model ranks among the top three, solving 69.8 % of problems at Mastery level. These numbers are noticeably lower than performance on high-resource Python/Java benchmarks of introductory level [3, 9], for two reasons: (1) our evaluation considers not only correctness but also output quality, whereas most prior studies focus solely on correctness; and (2) unlike standard function-level tasks, our benchmarks include *multi-task* problems in a low-resource language. To provide a baseline comparison with a domain-specific tool, we also evaluate BURST [35], a code synthesis system for OCaml, on the λCODEGEN benchmark. BURST produces correct solutions for only 11.3 % of the problems, performing significantly worse than general-purpose LLMs.

Note that our references to benchmarks in high-resource languages are indicative rather than direct comparisons. Differences in benchmark structure and model training make strict cross-language conclusions impossible. Instead, we use these references to illustrate a general trend: LLMs that perform strongly on popular, high-resource languages may exhibit lower performance on a low-resource language like OCaml [6, 61]. This framing allows readers to contextualize our findings without implying a formal cross-language comparison.

---

[1] https://www.digitaleducationcouncil.com/post/digital-education-council-global-ai-student-survey-2024. Accessed 2025-02-10.

Our results also reveal limitations: 40 % of tested LLMs achieve Mastery for less than 41 % of problems, with average performance below 50 %. All models struggle more with abstract concepts requiring theoretical implementation than basic programming tasks. Code generation from natural language proves more challenging than code repair, with LLMs performing 5 % better on syntax/type errors than on logical errors or full code generation. Further, we observe that performance of smaller models improves for code repair problems (about 20 % more Mastery level solution for Llama3.1 8B and Qwen2.5 7B). Overall, performance gaps among the models widens on λEXPLAIN, where GPT-4o solves 58.4 % of the problems, compared to Llama3.1 8B's 16.8 %. Taken together this reinforces the impression that data on which an LLM is trained on matters greatly and general-purpose LLMs struggle with more advanced and complex low-resource programming concepts.

These results carry implications for different audiences. For students, they highlight the importance of developing the skill to critically assess LLM outputs rather than relying on them uncritically—an ethical concern as much as a practical one. For instructors, our benchmarks provide a concrete way to raise awareness of both the strengths and shortcomings of LLMs and to design assessments that go beyond traditional problem solving, focusing instead on critique, debugging, and verification. For example, they can ask students to invent questions for the model to force wrong solutions [15]. For the programming languages (PL) community, our findings highlight both the adaptability of LLMs and the limitations of current domain-specific tools like BURST. This points to opportunities for integrating specification-aware reasoning into LLMs, combining their flexibility with synthesis techniques. More broadly, our benchmarks motivate the development of stronger synthesis and repair tools for low-resource languages and larger, more complex programs, where PL researchers can make unique contributions.

## 2 Methodology

In this section, we describe the programming problems and the conceptual questions from a 2nd year functional programming course at McGill University. The course is mostly taken by students in their 2nd or even 3rd and final year. We also describe our evaluation and grading policy of LLMs.

### 2.1 Benchmark λCODEGEN: Homework Programming Problems (10 Assignments)

For this evaluation, we draw on homework programming problems from the 2nd year functional programming course at McGill University used in Fall 2022. There are 10 assignments, containing 53 tasks in total. Table 1 presents a brief overview of the content for each assignment with the number of tokens. The average number

■ **Table 1** Homework Programming Problems from Fall 2022

| HW | # Tokens | # Tasks | Topics | Difficulty |
|---|---|---|---|---|
| 1 | 558 | 3 | Basics, recursion and pattern-matching | B |
| 2 | 993 | 8 | Basic recursive data types and tail-recursive functions | B |
| | | | More complex recursive data types and recursion | AP |
| 3 | 783 | 7 | Higher-order functions and arithmetic | B |
| | | | Higher-order folds to manipulate lists | AP |
| 4 | 1392 | 3 | Extract variable names from a boolean formula | B |
| | | | Exception-based backtracking and continuations | AP |
| 5 | 1074 | 3 | Find tree depth and subtrees using continuations | AP |
| | | | Continuation-passing style parser for arithmetic exp. | AP |
| 6 | 1292 | 13 | Option-based list traversal | B |
| | | | Using modules | B |
| | | | Converting between types and abstractions | AP |
| 7 | 409 | 2 | HOFs for lists and string manipulations | B |
| | | | Mutable state and records | B |
| 8 | 499 | 8 | Lazy programming and streams | B |
| | | | Streams and recursive sequence | AP |
| 9 | 3601 | 3 | Prog. lang. theory, type checking, interpreter | PT |
| 10 | 2748 | 3 | Prog. lang. theory, unification, type inference | PT |

of tokens per question is 257. The programming problems range from: 1) *Basic* techniques (labelled **B**) that include pattern matching on data types, basic higher-order functions (HOF), and recursion 2) *Advanced programming* techniques (labelled **AP**) such as backtracking problems, continuations, streams, and lazy programming; 3) *Programming Language Theory* which covers programming concepts including computing the free and unused variables, and extending an existing implementation of a type checker and interpreter (labelled **PT**).

## 2.2 Benchmark λREPAIR: Code with Syntax, Type, and Logical Errors (150 Problems)

Our second benchmark samples programs from actual buggy student code. These student programs were collected in Fall 2022 iteration of the functional programming course using a modified version of LearnOCaml platform [5, 19] which allows students to edit, compile, test, and debug OCaml code all in one place.

On top of the Learn-OCaml platform, we have built a data collection pipeline that automatically logs students' actions [8]. Specifically, we send *local* programming events like compile and evaluation (for testing and debugging) with asynchronous logging requests to our backend server. Figure 1 illustrates the process of collecting the data from the online environment Learn-OCaml.

Around 52.81 % (i.e., 169 out of 320) students gave us consent to analyze their interactions with the Learn-OCaml platform. Using the pipeline, we collected more than 270,000 programming events, each event stores a snapshot of the code as well as

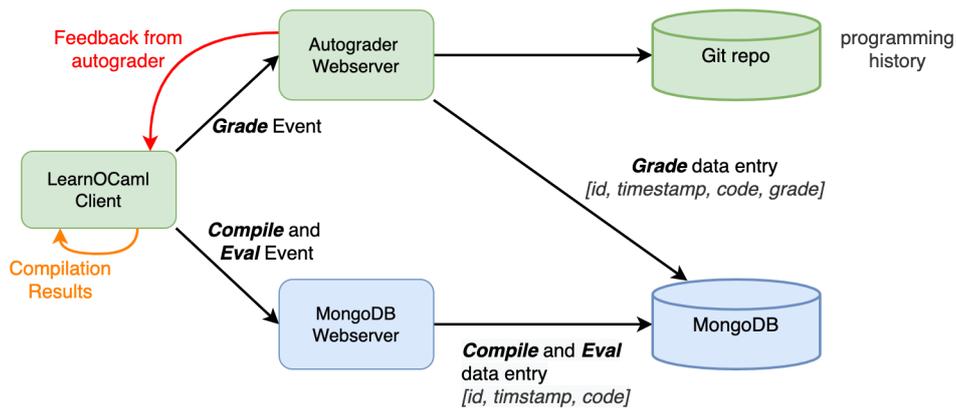

**■ Figure 1** Data collection pipeline. All submission data is stored in a *MongoDB* database. The components in light green are original components in the Learn-OCaml platform, while the components in light blue are newly introduced by us.

feedback information (e.g., static errors, grades, etc.). From these programming events, we sample 50 unique buggy programs for each of the three categories: syntax error, type error, and logical error. In contrast to *multi-function* evaluation in λCODEGEN, all buggy programs are *single-function* programs. We ensure that buggy programs are distributed evenly across 10 assignments to balance difficulty levels.

### 2.3 Benchmark λEXPLAIN: Conceptual Question (50 Problems)

This benchmark consists of 50 conceptual questions that instructor(s) used in the Fall 2024 and Fall 2021 iteration of this course. These questions include questions given to students as preparation for exams and questions that were used on actual exams. They test the understanding of variable scope, constructing induction proofs, deducing types, program evaluation, and applying substitution (i.e. replacing a free variable with an expression). These questions are to be answered without a compiler.

**■ Table 2** Conceptual questions grouped by topic and difficulty

| # Questions | Topic | Difficulty |
|---|---|---|
| 7 | Bindings, types, recursive functions | **B** |
| 10 | Bindings, types, recursive functions | **A** |
| 4 | Basic concepts in OCaml | **B** |
| 9 | Induction and reasoning about programs | **B** |
| 6 | Induction and reasoning about programs | **A** |
| 6 | Evaluation of programs | **B** |
| 8 | Evaluation of programs | **A** |

Each question is manually labelled with a difficulty level. In contrast to programming exercises, which are designed to build practical skills, conceptual questions typically focus on assessing a student's understanding of fundamental principles and theories, and they often require explanations, proofs, or high-level reasoning.

We labelled each question with a difficulty level, **B** (Basic) or **A** (Advanced). Table 2 shows the distribution of difficulty level across 50 questions with their topics.

## 2.4 Overview of Selected LLMs

For our evaluation of LLMs in functional programming, we selected models across several dimensions: leading systems from major providers (Meta, OpenAI, Google, Anthropic) to enable cross-company comparison; both closed-source paid models (Claude 3.7 Sonnet, GPT family, Gemini family) and free, open-source models (Llama 3 family, Qwen2.5 7B) to account for accessibility and cost, particularly relevant for students; models of different parameter sizes (e.g., 7B vs. 70B) from the same provider to assess the effect of scale; and both chat-oriented and reasoning-focused models to capture differences in problem-solving capability.

All LLM results were obtained through official API calls, without modifying sampling or decoding parameters. Unless otherwise noted, we used the default settings provided by each API (e.g., temperature, max tokens, top-p). While defaults are stable at the time of our experiments, future changes by providers may affect exact reproducibility.

Table 3 presents the 9 LLMs studied in our experiments with their company name, model sizes (Column "Size"[2]), and the model type.

## 2.5 Evaluation Methodology

We interact with all of the LLMs via OpenAI and OpenRouter APIs.[3] To ensure stability and reduce hallucination of LLMs, we prompt each model with the same benchmark problem 5 times. This results in 250 answers total (5 x 50 problems) for $\lambda$Repair and $\lambda$Explain and 50 answers total for $\lambda$CodeGen (5 x 10 assignments). We use both automated and manual grading to assess the responses generated by the LLM.

Our evaluation methodology is guided by two principles: first, we prompt the LLMs in the way we expect students would use them, and second, we evaluate the responses from the perspective of an instructor, using criteria such as correctness, algorithm design, and readability. For homework assignments, we ask for *code-only* output and provide the model with an entire homework containing multiple programming tasks, which makes code generation more challenging than the commonly used single-task evaluations. This allows us to better gauge the usefulness and effectiveness of LLMs in an educational setting. We ask for a *concise* answer or explanation in the prompt to avoid verbosity. To support reproducibility and reuse, the benchmark, along with prompting scripts and instructions, will be provided as an artifact [60]. The package will be well-structured and documented, enabling other researchers to deploy it easily and apply it to new models.

---

[2] Parameter counts for several models are not officially disclosed by their providers. We report approximate magnitudes from publicly discussed estimates: GPT-4o-mini ≈ 8B, GPT-4o ≈ 200B, Claude Sonnet 3.7 ≈ 400B, Gemini 2.0 Flash ≈ 200 - 300B, and o3-mini — unknown. These values are anecdotal or speculative and should be interpreted as rough estimates; entries marked with "–" indicate no reliable public information at the time of writing.

[3] https://openrouter.ai/. Accessed 2025-02-10.

■ **Table 3** State-of-the-art LLMs considered in our evaluation, listed by company, parameter size, and model type.

| Model | Company | Size | Type |
|---|---|---|---|
| Claude 3.7 Sonnet [4] | Anthropic | – | Chat/Reasoning |
| GPT-4o [5] | OpenAI | – | Chat/Reasoning |
| GPT-4o Mini | OpenAI | – | Chat/Reasoning |
| o3-mini | OpenAI | – | Chat/Reasoning |
| Llama 3.1 70B Instruct [6] | Meta | 70B | Chat |
| Llama 3.1 8B Instruct | Meta | 8B | Chat |
| Gemini 2.0 Flash [7] | Google DeepMind | – | Chat/Reasoning |
| Gemini 1.5 Flash 8B | Google DeepMind | 8B | Chat/Reasoning |
| Qwen2.5 7B [55] | Alibaba Cloud | 7B | Base (General-purpose) |

**Set-up and Assessment of Benchmark $\lambda$CODEGEN** For $\lambda$CODEGEN we give the entire natural language homework description together with the type specifications of a given task. We then first grade each generated answer using the compiler and auto-grader (automated assessment) followed by a manual assessment of the 50 generated answers which is done by 2 experienced teaching assistants. During the manual assessment, each solution is checked whether the generated code meets problem requirements. Typical issues include: 1) Not providing a tail-recursive function or not using specific HOFs, although the problem requires it; 2) Using disallowed functions or imperative features where the problem description forbids the use of library functions and / or state; 3) Providing an overall inefficient algorithms. Such manual assessment is usually absent in current studies (e.g. [22]).

We grade answers with respect to three categories: correctness, algorithm design, and readability (see Table 4). In each of these categories, a generated answer is then ranked as mastery, proficient, developing, beginner, or non-gradable. For a solution to be ranked at the mastery level, it must be correct (i.e. pass all test-cases from the auto-grader), and follow correctly the problem requirement. If it exhibits any of the aforementioned three issues, it will be classified as Proficient or lower.

We grade answers following a hierarchical dependency, ensuring that Correctness(Understanding of the problem) > Algorithm Design (Strategy) > Readability—i.e., correctness is a prerequisite for evaluating algorithm design, and algorithm design is a prerequisite for readability. If a solution fails to compile, it cannot be auto-graded and it is deemed incorrect and receives zero. Subsequent criteria, algorithm design and readability, then become irrelevant. The final grade for each generated answer is the average of its scores across the three criteria.

---

[4] https://www.anthropic.com/news/claude-3-7-sonnet. Accessed 2025-02-10.

[5] https://openai.com/index/hello-gpt-4o/. Accessed 2025-02-10.

[6] https://www.llama.com/models/llama-3/. Accessed 2025-02-10.

[7] https://deepmind.google/models/gemini/flash/. Accessed 2025-02-10.

■ **Table 4** Grading rubric for code generation. Solutions containing syntax or type errors will be labelled as Non-gradable.

| Mastery (4 points) | Proficient (3 points) | Developing (2 points) | Beginning (1 point) |
|---|---|---|---|
| **Correctness** | | | |
| The solution passes all test cases and has no style issues. | The solution passes at least 75 % of test cases and has no style issues. | The solution fails many test cases (between 75 % and 25 % pass rate). | The solution compiles but passes fewer than 25 % of the test cases. |
| **Algorithm Design** | | | |
| The algorithm or implementation techniques match the problem specification. | The algorithm or implementation techniques mostly match the problem specification. | The algorithm or implementation techniques barely match the problem specification. | The algorithm or implementation technique is incoherent or inappropriate (e.g., uses disallowed or unnecessary helper functions). |
| **Readability** | | | |
| Generates only the required code. | Generates code but includes redundant functions or test cases. | Generates code but includes excessively verbose or unnecessary natural language explanations. | N/A (if the solution is ranked as Beginning with respect to correctness and algorithm design, it is also ranked as Beginning with respect to readability). |

**Set-up and Assessment of Benchmark $\lambda$Repair**  For $\lambda$Repair, we give to the LLM a buggy program along with the corresponding error message(s) and ask the LLM to generate a repaired version without prior examples, i.e. we prompt the model in a *zero-shot* setting. Note that we test the model using single-error buggy programs since our compiler only identifies the *first* error in the program. For assessing answers for $\lambda$RepairSyntax and $\lambda$RepairType, we give the syntax / type error and error messages generated by the OCaml compiler to the LLM together with the buggy program. For assessing answers $\lambda$RepairProg, we give the buggy program together with the *single-function* homework description (i.e. the question of the homework that corresponds to this single-function) to the LLM.

We omit any test cases that were expected, passed, or missed. We grade the generated answers using automated and manual assessment, in the same way as the $\lambda$CodeGen problems (see Table 5).

**Set-up and Assessment of Benchmark $\lambda$Explain**  Last, for $\lambda$Explain we prompt the model with natural-language exam question. For questions requiring a structured response (e.g., a state diagram), we provide an example to guide the model, ensuring that its lack of prior formatting knowledge does not affect performance. We adjusted the grading scale for programming problems to be better suited for conceptual problems in Table 6. In particular, we take into account the understanding of the problem, strategy used of solving the problem, and communication of the explanation. Our

■ **Table 5** Grading rubric for code repair. Responses containing unchanged code, non-OCaml code, garbage text, or no code will be rated as Non-gradable.

| Mastery (4 points) | Proficient (3 points) | Developing (2 points) | Beginning (1 point) |
|---|---|---|---|
| **Correctness** | | | |
| Repairs code perfectly (i.e., code that passes the compiler). | Repairs code so that it passes the compiler, but makes redundant changes (e.g., adds additional helper functions or test cases) or reformats code unnecessarily. | Correctly localizes the error, but fails to repair the code or introduces a new error of the same type. | Fails to localize the error or repair the program. |

■ **Table 6** Grading rubric for conceptual / non-programming questions. Off-topic solutions or those containing garbage text will be labelled as Non-gradable.

| Mastery (4 points) | Proficient (3 points) | Developing (2 points) | Beginning (1 point) |
|---|---|---|---|
| **Understanding of the problem** | | | |
| Demonstrates a deep understanding of the problem, identifying all key components and potential challenges. | Shows a good understanding of the problem, identifying most key components and challenges. | Shows a basic understanding of the problem but may miss some key components or misunderstand certain aspects. | Struggles to understand the problem and its requirements, leading to significant inaccuracies in the solution. |
| **Strategy** | | | |
| Articulates a clear strategy to correctly solve the problem. | Articulates only parts of an appropriate strategy, but most key elements are present. | Articulates a strategy that is only partially useful; key elements might be missing or inappropriate steps are provided. | Fails to provide a strategy to address the problem. |
| **Readability** | | | |
| Provides a clear, complete, unambiguous, and concise explanation. | Provides a complete explanation but may lack clarity or conciseness in some areas. | Provides an unclear, inconsistent, or incomplete explanation; contains unwanted information. | Provides an incoherent explanation or fails to give one. |

grading again adapts a hierarchical dependency: understanding > strategy > communication. A response that lacks a correct understanding cannot have a meaningful strategy, and a weak strategy undermines effective communication.

## 2.6 Overall Letter Grade of LLM

Using the mastery grading approach described in the previous section, we assign to each solution generated by an LLM a Mastery, Proficient, Developing, Beginning, or Non-gradable level. For each LLM, we then compute the overall percentage of problems solved at a given level: #(problems at this level) / #(total questions). To better classify LLMs we also compute a letter grade for a LLM by converting the mastery levels to letter grades using university's grade conversion policy (shown in Table 7). The overall grade of an LLM is the average over all grades. For example, if an LLM has: 80 % Mastery , 10 % Proficiency, and 10 % Developing, its final letter grade is an A- (8.0).

■ **Table 7** Grading conversion policy for computing the letter grade of LLM.

| Mastery Level | Letter Grade | Mid-point Grade | Grade Range |
|---|---|---|---|
| Mastery | A | 8.45 | 8.2-8.7 |
| | A- | 7.95 | 7.7-8.2 |
| | B+ | 7.45 | 7.2-7.7 |
| Proficiency | B | 6.95 | 6.7-7.2 |
| | B- | 6.45 | 6.2-6.7 |
| | C+ | 5.95 | 5.7-6.2 |
| Developing | C | 5.45 | 5.2-5.7 |
| Beginning | D | 4.95 | 4.7-5.2 |
| N/A | F | 0.0 | 0.0 |

## 3 Results

We now revisit our research questions from the introduction.

For all of these questions, our evaluation results reveal clear performance tiers across all tasks. The models o3-mini, Claude 3.7 Sonnet, and GPT-4o consistently emerge as the top three LLMs. Conversely, Gemini 1.5 Flash 8B, Qwen2.5 7B, Llama3.1 8B, and occasionally GPT-4o-mini consistently rank at the bottom. We therefore group the ten models into three tiers in tables for easier comparison.

### 3.1 RQ1: Are LLMs Solving Functional Programming Assignments Correctly, and Concisely?

Our evaluation of nine LLMs on λCODEGEN (Table 8) reveals a significant performance gap among them, i.e. the standard deviation for the weighted total grade / Mastery rate (%) is 15.8/ 20.6 respectively. The top three of models – o3-mini, Claude 3.7 Sonnet , and GPT-4o – demonstrate strong capabilities with Mastery rates around 70 %, achieving a weighted overall grade in the B range. However, smaller closed-resource models like Qwen2.5 7B and Llama3.1 8B show substantially lower performance at 21.9 % and 15.8 % respectively, resulting in a F grade. Apart from the top models, 40 % of the LLMs (the bottom 4 LLMs) in our study were only able to generate mastery level solutions for less than 41 % of the problems. On average the LLMs were able to solve about half (52.6 %) of the homework programming problems at the mastery level (see Table 8). This is still a sobering result.

On λCODEGEN, the top three models ( GPT-4o, o3-mini and Claude 3.7 Sonnet) are quite effective since they achieve Mastery-level solutions for 69.8 %, 74 % and 71.3 % of problems, respectively. These success rates are lower than those reported on high-resource Python/Java benchmarks such as HumanEval [9] and MBPP [3], where

■ **Table 8** Evaluation of code generated for benchmark λCODEGEN
(M := Mastery; P := Proficient; D := Developing; B := Beginning; N := Non-gradable)
Rows highlighting with green are results from commercial (paid) models whereas the rest
are from open-source (free) models

| Model | M (%) | P (%) | D (%) | B(%) | N (%) | Grade | Avg. Tier Grade |
|---|---|---|---|---|---|---|---|
| o3-mini | 73.6 | 3.4 | 2.3 | 12.1 | 8.7 | B+ | |
| Claude 3.7 Sonnet | 71.3 | 3.8 | 4.5 | 8.3 | 12.1 | B | B |
| GPT-4o | 69.8 | 3.0 | 3.0 | 12.5 | 11.7 | B | |
| Gemini 2.0 Flash | 58.9 | 1.5 | 2.6 | 9.1 | 27.9 | C+ | C+ |
| Llama3.3 70B | 58.5 | 2.3 | 1.9 | 10.9 | 30.9 | C+ | |
| Gemini 1.5 Flash 8B | 40.8 | 0.8 | 2.6 | 13.2 | 42.6 | D | |
| GPT-4o-mini | 40.4 | 0.0 | 6.4 | 24.9 | 28.3 | D | F |
| Qwen2.5 7B | 21.9 | 0.8 | 4.2 | 22.3 | 50.9 | F | |
| Llama3.1 8B | 15.8 | 0.0 | 4.2 | 12.1 | 67.9 | F | |

| Avg. Grade | Median Grade | Grade SD | Avg. M (%) | Median M (%) | SD M (%) |
|---|---|---|---|---|---|
| 54.70 | 58.0 | 15.8 | 52.6 | 58.7 | 20.6 |

top models exceed 90 % accuracy. Nonetheless, LLMs remain effective on λCODEGEN compared to large real-world programs like SWE-Bench [22]. Compared to earlier generations on low-resource languages, our results outperform Codex in 2022 (about 40 % pass rate) [7] and Llama using knowledge transfer [6], indicating that general-purpose LLMs nowadays can achieve strong results in low-resource languages and that continued model evolution may further narrow the gap with high-resource settings.

Compared to specialized code synthesis tools [35, 39], LLMs offer different tradeoffs: domain specific tools require precise inputs and type signatures and guarantee correct solutions for small FP tasks (15–30 AST nodes), whereas LLMs can tackle larger, more open-ended tasks at the cost of occasional hallucinations. Extending this comparison, we run BURST [35], a state-of-the-art OCaml synthesizer based on bottom-up search with angelic execution, on λCODEGEN in a single-function setting.

Our results show that BURST is only able to synthesis 11.3 % of the problems, primarily those involving structural recursion such as tree traversal operations. Although BURST demonstrates particular strength in reliably generating correct type signatures and implementations for recursive functions with precise specifications, its limitations are apparent on CPS (continuation-passing style), higher-order functions, lazy or infinite streams, memorization, exception-based backtracking, and multi-step reasoning tasks. However, when it succeeds, it reliably generates correct type signatures, and performs well on simple recursion functions (e.g. tree traversal). This contrast illustrates the respective roles of both approaches: specialized tools provide guaranteed correctness within narrower, well-defined domains, while LLMs can adaptively generate code from natural-language prompts, explaining why novices may choose to use LLMs over domain-specific tools.

The four LLMs ranked the lowest generate often code classified as "Non-gradable" where the generated code does not even compile (Llama3.1 8B at 67.9 %, Qwen2.5 7B at 50.9 %, and Gemini 1.5 Flash 8B at 42.6 %). The most common causes for failing to compile are syntax and type errors in addition to invalid code structures (6.1 % of Llama 3.1 8B responses) and incorrect comment formats (7.5 % of Qwen2.5 7B responses). The Llama family models uniquely produced some garbage text as response.

We also note that across all LLMs, the percentage of responses classified as beginner is higher than the percentage of responses classified as proficient or developing. Our manual assessment reveals that responses classified as beginner typically type-check, but do not satisfy the given specification, contain logical errors, or pass few or no test cases. Even top models such as GPT-4o occasionally fails to adhere to prompts for example using HOFs in homework 7 where they were explicitly disallowed; or vice versa when o3-mini created unnecessary helper functions in homework 3 instead of directly employing required HOFs. This seems to suggest that LLMs do not necessarily understand the full specification given. While GPT-4o is among the top LLMs for functional programming assignments, its Mastery rate of 69.8% reveals significant room for improvement. These results suggest that overall LLMs are not yet reliable enough to effectively solve functional programming tasks but provide great adaptivity for novices.

## 3.2 RQ2: Are LLMs Repairing Syntax, Type, and Logical Errors in Functional Programs Correctly and Concisely?

We assess here how helpful LLMs can be in error repair – from syntax and type errors to logical errors – in actual student programs.

**RQ2a: Are LLMs Repairing Syntax Errors Correctly and Concisely?** The results in Table 9 show that for syntax error repair the top-three LLMs ( o3-mini, GPT-4o, and Claude 3.7 Sonnet) can achieve over 78% at Mastery level in our examples. Across all LLMs we observe a clear tiered decline: GPT-4o-mini (67%), and Gemini 2.0 Flash (62%) form a middle tier, while open-source models (Llama3.3 70B, Qwen2.5 7B) achieve 42-59% Mastery. Notably, most LLMs (6 out of 10) successfully repair at least 60% of buggy programs. For lower-tier models, we also observe a notable improvement in their performance on code repair compared to code generation.

■ **Table 9** Evaluation of code generated for benchmark $\lambda$REPAIRSYNTAX
(M := Mastery; P := Proficient; D := Developing; B := Beginning; N := Non-gradable)
Rows highlighting with green are results from commercial (paid) models whereas the rest are from open-source (free) models

| Model | M (%) | P (%) | D (%) | B (%) | N (%) | Grade | Avg. Tier Grade |
|-------|-------|-------|-------|-------|-------|-------|-----------------|
| o3-mini | 82.4 | 3.2 | 3.2 | 1.6 | 9.6 | A- | |
| GPT-4o | 81.2 | 8.4 | 6.4 | 4.0 | 0.0 | A- | |
| Claude 3.7 Sonnet | 78.4 | 5.2 | 14.0 | 2.4 | 0.0 | A- | A- |
| GPT-4o-mini | 67.2 | 2.4 | 24.4 | 6.0 | 0.0 | B+ | |
| Gemini 2.0 Flash | 62.0 | 8.4 | 19.2 | 6.4 | 4.0 | B+ | B+ |
| Llama3.3 70B | 58.8 | 2.4 | 24.4 | 6.4 | 8.0 | B | |
| Qwen2.5 7B | 47.6 | 2.4 | 44.4 | 2.4 | 3.2 | B | |
| Gemini 1.5 Flash 8B | 43.6 | 5.2 | 40.0 | 5.4 | 5.8 | B | B |
| Llama3.1 8B | 42.4 | 8.4 | 30.4 | 5.6 | 13.2 | B- | |

| Avg. Grade | Median Grade | Grade SD | Avg. M (%) | Median M (%) | SD M (%) |
|------------|--------------|----------|------------|--------------|----------|
| B+ | B+ | 4.96 | 64.7 | 64.6 | 15.6 |

The top-performing LLMs surpass or are on par with specialized tools like SYN-SHINE's which achieves 82% accuracy on small buggy Java programs (under 100 tokens) [1]. Furthermore, RING, which is a syntax error repair tool powered by Codex and few-shot learning [23] reached 94 % accuracy on Python. This suggests that results could improve with example-based prompting, motivating our one-shot learning in later section.

Overall, our results further confirm that LLMs are becoming competitive with specialized syntax-repair tools while offering greater flexibility.

**RQ2b:Are LLMs Repairing Type Errors Correctly and Concisely?** While syntax error correction showed strong results, type errors present a more challenging scenario (Table 10). The top three commercial models ( o3-mini, GPT-4o, Claude 3.7 Sonnet) still achieve impressive Mastery rates (72-83 %). However, performance of open-source models (Llama3.3 70B: 60.4 %, Llama3.1 8B: 40.8 %), drops off more sharply compared to syntax error repair (λRεραιrSyntax benchmark suite). This suggests that these LLMs have not been trained on reasoning problems.

■ **Table 10** Evaluation of code generated for benchmark λRεραιrType
(M := Mastery; P := Proficient; D := Developing; B := Beginning; N := Non-gradable)
Rows highlighting with green are results from commercial (paid) models whereas the rest are from open-source (free) models

| Model | M (%) | P (%) | D (%) | B (%) | N (%) | Grade | Avg. Tier Grade |
|---|---|---|---|---|---|---|---|
| o3-mini | 80.2 | 5.6 | 11.2 | 3.0 | 0.0 | A- | |
| GPT-4o | 75.6 | 3.2 | 16.4 | 6.2 | 0.0 | A- | |
| Claude 3.7 Sonnet | 72.0 | 2.4 | 20.4 | 5.2 | 0.0 | A- | A- |
| Llama3.3 70B | 60.4 | 1.6 | 18.6 | 11.4 | 8.0 | B+ | |
| GPT-4o-mini | 57.6 | 3.0 | 26.4 | 11.0 | 2.0 | B+ | B+ |
| Gemini 2.0 Flash | 55.6 | 4.0 | 22.4 | 12.0 | 6.0 | B+ | |
| Qwen2.5 7B | 42.4 | 4.0 | 36.4 | 13.0 | 4.0 | B | |
| Llama3.1 8B | 40.8 | 3.0 | 25.0 | 17.2 | 14.0 | B | B |
| Gemini 1.5 Flash 8B | 37.6 | 2.0 | 41.4 | 14.0 | 5.0 | B- | |

| Avg. Grade | Median Grade | Grade SD | Avg. M (%) | Median M (%) | SD M(%) |
|---|---|---|---|---|---|
| B+ | B+ | 4.96 | 60.5 | 59.0 | 15.9 |

Comparing LLMs to specialized type error correction tools, there is no clear winner. On the one hand, LLMs outperform OCaml-specific systems like RITE, which also uses students buggy code , and MENTAT [44, 45], which report only 39 %/45 % success on large-scale OCaml benchmarks. However, those systems could be more efficient on smaller programs like our benchmark, which motivates a valuable comparison for future work. On the other hand, LLMs fall short of language-specific tools such as PyTy for Python, which achieves 85.4 % accuracy [10]. Recent work on Haskell [46] shows a complementary approach: combining GPT-4 with compiler-based fault localization and QuickCheck validation yields 88.5 % repair success.

Overall, these results indicate that general-purpose LLMs may not match the precision of specialized type error correction tools developed for a particular language. At the same time, LLMs demonstrate surprisingly strong cross-language capabilities, even for OCaml's complex type system where few dedicated tools exist, making them a potentially valuable resource for beginners learning a new language. LLMs perform particularly well on single-function syntax and type error repair, likely because these tasks isolate the target error and require only localized fixes, unlike multi-task code generation which requires complex reasoning ability.

**RQ2c:Are LLMs Repairing Logical Errors Correctly and Concisely?**   The top three LLMs (o3-mini, Claude 3.7 Sonnet, and GPT-4o) demonstrate mastery rates of above 60 % (Table 11). While this shows that their capabilities in logical error repair are weaker than syntax or type error repair, these models consistently produce fully correct solutions. The mid-tier includes models such as GPT-4o-mini and Llama3.3 70B, which show moderate success but a higher proportion of partially correct or buggy responses. The bottom tier (Gemini 1.5 Flash and Qwen2.5) performs significantly worse with Mastery rate lower than 45 %.

■ **Table 11**   Evaluation of code generated for benchmark $\lambda$REPAIRPROG. (M := Mastery;  P := Proficient;  D := Developing;  N := Non-gradable)
Rows highlighting with green are results from commercial (paid) models whereas the rest are from open-source (free) models

| Model | M (%) | P (%) | D (%) | B (%) | N (%) | Grade | Avg. Tier Grade |
|---|---|---|---|---|---|---|---|
| o3-mini | 72.4 | 5.1 | 6.0 | 6.5 | 10.0 | A- | |
| Claude 3.7 Sonnet | 68.0 | 6.0 | 3.0 | 1.0 | 15.0 | A- | |
| GPT-4o | 67.2 | 3.6 | 1.6 | 7.0 | 20.6 | A- | A- |
| Llama3.3 70B | 56.2 | 1.6 | 5.2 | 16.0 | 21.0 | C+ | |
| GPT-4o-mini | 52.8 | 4.0 | 11.0 | 12.2 | 20.0 | B+ | B |
| Gemini 2.0 Flash | 48.8 | 3.6 | 8.0 | 18.6 | 21.0 | B | |
| Qwen2.5 7B | 43.4 | 4.0 | 5.0 | 19.6 | 24.0 | C | |
| Llama3.1 8B | 37.6 | 3.0 | 7.0 | 29.4 | 20.0 | B | B- |
| Gemini 1.5 Flash 8B | 33.6 | 3.0 | 5.0 | 39.0 | 24.0 | B | |

| Avg. Grade | Median Grade | Grade SD | Avg. M (%) | Median M (%) | SD M (%) |
|---|---|---|---|---|---|
| B+ | B | 6.52 | 55.5 | 54.5 | 14.31 |

Previous APR(Automated Program Repair) tools for education, such as CLARA and Refactory ([18, 21]) for Python, also rely on existing correct submissions or test cases to repair student buggy assignment. These systems are designed to produce scalable, real-time feedback in educational settings. While Refactory and CLARA report over 90 % repair success on their respective Python benchmarks, the top-tier LLMs can achieve a reasonably high Mastery rates while being easy to (re)use across different languages such as OCaml.

**Further Investigation on Code Fixing Using One-shot Learning**   In addition to the zero-shot evaluation, we assess one-shot learning for $\lambda$REPAIRPROG for syntax and type errors. One-shot learning refers to a setup where the model is given one example (one demonstration) of a task, and then asked to perform that task on new inputs. It differs from few-shot learning, where multiple examples are shown, and from zero-shot which shows none.

This setup allows us to assess whether minimal in-context guidance improves code-repair performance. We focus on GPT-4o, widely used by students, and Llama 70B, an open-source model. As shown in Table 12, one-shot learning consistently improves performance over zero-shot prompting for both models although the overall gain is small. GPT-4o gains 4.4 % and 2.2 % in Mastery for syntax and type errors, respectively, with fewer beginner-level responses. Llama 70B shows slightly larger gains and reduces non-gradable outputs by up to 6.6 %, aligning with the intuition that weaker models benefit more from contextual guidance. Given the relatively small sample size of 50 buggy programs, these results should be seen as suggestive trends rather than definitive improvements, but they point toward the potential of even minimal in-context examples to steer LLMs toward better repair performance.

■ **Table 12**  Performance improvement (percentage point changes) from zero-shot to one-shot prompting for GPT-4o and Llama-70B. 0.0 here mean no change.

| Error Type | Model | Δ Mastery | Δ Proficient | Δ Beginning | Δ Non-gradable |
|---|---|---|---|---|---|
| Syntax | GPT-4o | 4.4 | 2.0 | −6.4 | 0.0 |
| | Llama-70B | 5.6 | 3.8 | −2.8 | −6.6 |
| Type | GPT-4o | 2.2 | 2.0 | −4.2 | 0.0 |
| | Llama-70B | 3.6 | 4.0 | −2.0 | −5.6 |

## 3.3 RQ4: Do the LLMs' Abilities to Generate Correct and Concise Answers Vary Across Different Types of Questions—From Code Generation vs. to Code Repair vs. to Explaining Theoretical Concepts?

■ **Table 13**  Evaluation of long-answers for λExplain.
(M := Mastery;  P := Proficient;  D := Developing;  B := Beginning;  N := Non-gradable)
Rows highlighting with green are results from commercial (paid) models whereas the rest are from open-source (free) models

| Model | M (%) | P (%) | D (%) | B (%) | N (%) | Grade | Avg. Tier Grade |
|---|---|---|---|---|---|---|---|
| o3-mini | 80.0 | 15.2 | 4.8 | 0.0 | 0.0 | A | |
| Claude 3.7 Sonnet | 60.0 | 18.4 | 22.8 | 0.4 | 0.4 | A- | A- |
| GPT-4o | 58.4 | 13.6 | 23.2 | 4.8 | 0.0 | B+ | |
| Gemini 2.0 Flash | 44.4 | 10.8 | 21.6 | 15.2 | 8.0 | B+ | |
| GPT-4o-mini | 48.8 | 9.2 | 20.4 | 20.0 | 1.6 | B+ | B |
| Llama3.3 70B | 42.8 | 2.4 | 28.4 | 12.4 | 14.0 | B- | |
| Gemini 1.5 Flash 8B | 36.8 | 16.8 | 30.4 | 13.6 | 2.4 | B | |
| Qwen2.5 7B | 34.8 | 21.6 | 22.4 | 17.6 | 3.6 | B- | B- |
| Llama3.1 8B | 16.8 | 9.6 | 12.4 | 49.2 | 12.0 | B- | |

| Avg. Grade | Median Grade | Grade SD | Avg. M (%) | Median M (%) | SD M (%) |
|---|---|---|---|---|---|
| C | C+ | 15.8 | 50.4 | 46.6 | 19.1 |

Last, we assess models on there capability to explain fundamental concepts in programming language design. From a student perspective this is a particularly important area as it helps students to study the material and obtain additional explanations beyond what the instructor, teaching assistants, or course notes provide. It is also an area where it is more difficult for students to assess the correctness, as there is often no easy automated way that can be employed to test the generated response.

The results in Table 13 show large differences among the 9 evaluated LLMs with respect to their capability of generating explanations and responses to questions from the λExplain benchmark. We note that o3-mini continues to excel achieving an overall A grade. Its Mastery rates further demonstrate their ability to provide correct and concise explanations over 80 % of the time.

However, the majority of models struggle. The average weighted grades for mid-tier and lower-tier models are B and B- respectively, suggesting a significant gap in LLMs' ability to explain theoretical concepts. Compared to $\lambda$CODEGEN, the overall performance on $\lambda$EXPLAIN appears stronger primarily due to significantly lower Non-gradable rates. However, this is misleading as many responses still contain inaccuracies or unclear explanations as shown in Proficient, Developing ,and Beginning percentages. This is particularly problematic, as students are usually unable to verify or test the generated response. Our results align with existing research, including [24] who observe that over half of ChatGPT's StackOverflow answers contained errors. Similarly, Sarsa, Denny, Hellas, and Leinonen [47] find that Codex produced 32.8 % of the time incorrect explanations.

A common issue is unnecessary verbosity, although we explicitly ask for concise explanations in the prompt. For example, when asked to provide a tail-recursive OCaml program, models like Claude Sonnet and Gemini Flash often supply both tail-recursive and non-tail-recursive versions, followed by an long explanation of the importance of tail recursion in general. These responses are graded as Proficient, although they answer the main question correctly. However the verbosity of the answer distracts from the essence of the answer and the generated answer is slightly off topic.

### 3.4 RQ4: Do the LLMs' Abilities to Generate Correct and Concise Answers Vary across Different Types of Questions – From Code Generation vs. to Code Repair vs. to Explaining Theoretical Concepts?

To answer this question we compare the overall letter grade that we assigned to each LLM for a given task (see Figure 2). This demonstrates a clear differences in how well LLMs perform across the three tasks. We observe that all LLMs consistently struggle the most for code generation problems ( $\lambda$CODEGEN). For this benchmark, even the best models only get B grades, while smaller free models like Gemini 1.5 Flash 8B and Llama 3.1 8B score much worse (D or below). Our comparison also shows that LLMs have different strengths. While models such as o1, o3-mini but also Gemini 2.0 Flash and surprisingly Llama 3.1 8B score the highest on the $\lambda$EXPLAIN benchmark, all other LLMs score highest on problems from $\lambda$REPAIRSYNTAX. Furthermore, we observe a very large gap of 3-4 grades between their ability to generate code ( $\lambda$CODEGEN) and the other tasks for several models: Qwen 2.5 7B, Llama 3.1 8B, GPT-4o mini, and Gemini 1.5 Flash 8B. For top performing models this performance gap between different tasks is much more narrow.

### 3.5 RQ5: How Does the Difficulty Level of a Question Impact the LLMs Response across Different Types of Questions?

We also compute the Mastery percentages on basic/advanced questions by counting the number of problems in each level and divide by total basic/advanced problems in $\lambda$CODEGEN and $\lambda$EXPLAIN. We present the results in Figure 3. For code generation,

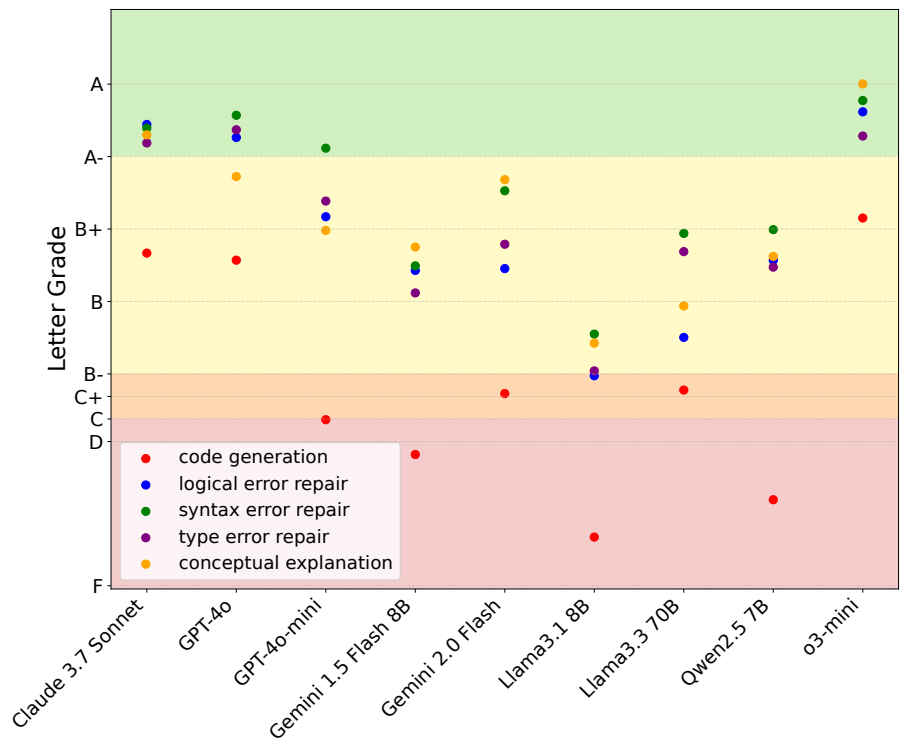

■ **Figure 2**  Weighted grade comparison of LLMs across different tasks – from code generation, to code repair and to explaining theoretical concepts we observe that all LLMs generally perform better on basic questions compared to advanced programming problems, although the difference is not very large. Top-performing models (Claude 3.7 Sonnet, GPT-4o, Gemini 2.0 Flash, and o3-mini) achieve Mastery rates above 60 % for both difficulty levels. This shows they can handle not just simple tasks but more advanced programming techniques reasonably well.

Basic tasks, such as pattern matching, simple tail recursion, and higher-order functions, are solved at higher rates, particularly when the problems are short and clearly defined. Advanced tasks, including recursion, continuations, lazy programming, or larger code generation problems, result in more type errors, suggesting that LLMs can generate correct code for simpler tasks but lack deeper type-based reasoning for complex OCaml programs.

The gap is more pronounced for programming theory (**PT**) questions. Mastery rates for basic questions are nearly two to three times higher than for **PT** problems, with weaker models like Llama 3.1 8B and Qwen2.5 7B scoring 0 % Mastery. These theory questions involve implementing parsers and a programming language similar to OCaml, and test concepts such as free variables, substitution, evaluation, and type inference. The poor performance highlights a key limitation: while LLMs can generate working code for programming problems, they struggle with tasks requiring deep understanding of programming language theory and precise type reasoning. This is as expected since current models may be pattern-matching from training examples rather than truly understanding underlying computational theories.

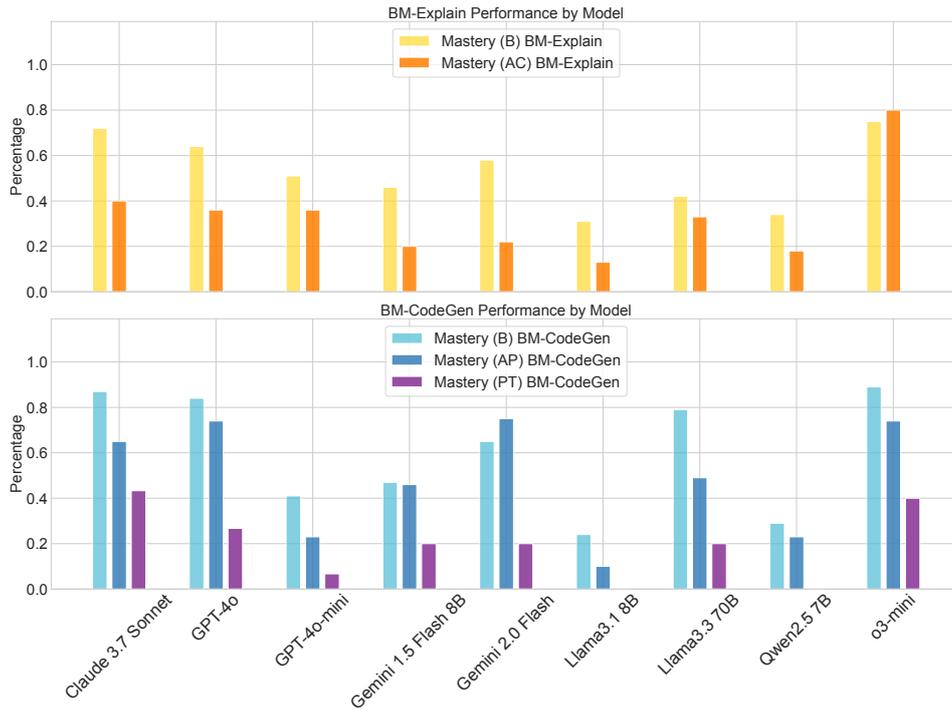

■ **Figure 3** Mastery rates for **B**(Basic), **AC**(Advanced Concept)/**AP**(Advanced Programming) and **PT**(Programming Technique) problems across models. BM-CODEGEN stands for $\lambda$CODEGEN whereas BM-Explain stands for $\lambda$EXPLAIN.

In answer generation ($\lambda$EXPLAIN), difficulty trends are similar: basic questions are easier, with Mastery rates for advanced (**AC**) questions roughly half that of basic (**B**) questions for most models. Top-tier reasoning models, including GPT-4o and o3-mini, achieve higher Mastery on **AC** questions but still struggle with complex topics such as type inference, substitution, and dynamic evaluation. For example, they often produce correct intermediate reasoning but fail to combine steps into accurate final answers. Variable scope issues (overshadowing and overwriting) frequently lead to incorrect explanations and examples. In contrast, these models handle tail-recursive analysis well although there are style issues in the code implementation, which means they may have the reasoning ability to analyze tail-recursive functions but struggle implementing the concept in code.

## 4    Threats to Validity

One limitation of this study is the evaluation process which could introduce biases. For example, the grading of generated responses into Mastery, Proficient, Developing, Beginning, and Non-gradable categories is performed manually, which could be subjective. Incorporating multiple graders could help mitigate this concern and improve the reliability of the classification. In the $\lambda$REPAIR benchmark, we assess whether a model resolves the original error type (syntax, type, or logical) but do not penalize new errors introduced during repair. This follows standard practice in repair tools, which target specific error types, allowing performance comparisons across error categories. Nevertheless, tracking both initial and final correctness could provide a more comprehensive measure of model effectiveness. Also, running λREPAIR on a competing repair tool is also another valuable direction of future work. Furthermore, LLMs are provided with the original problem description along with the buggy code in fixing logical errors. However, it remains unclear whether a model genuinely attempts to repair the given logical error or simply generates a new solution to the problem from scratch. While one could test this by supplying only test cases, our setup was designed to mimic the student experience, where such test cases are typically unavailable. This limitation highlights an opportunity for deeper investigation and is noted as a direction for future work.

Another consideration is potential overlap between our evaluation data and the models' training sets. Since our data is from Fall 2022 they could have been seen during training. While this limitation is common in studies using publicly available or classroom-derived data, performance evaluation remains meaningful, as LLMs can still fail on familiar tasks. Collecting newer datasets would require renewed student consent and is left for future work. We also assume that LLMs are fully trained and do not undergo additional learning during evaluation, which is critical for validity and reproducibility. In addition, the cost and accessibility of large-scale LLM evaluations introduce equity concerns, as not all researchers or educators may have the resources to replicate this study. For the same reason, our evaluation does not include all available models: we exclude the o1 model due to its high cost, which makes large-scale reproduction impractical. Preliminary experiments indicated performance comparable to o3, which we retain as a more affordable representative. Consequently, our results should be interpreted as indicative rather than exhaustive.

## 5 Related work

We mainly introduce related work on LLMs for software engineering and LLM evaluation in this section. In the field of software engineering (SE), LLMs have shown remarkable potential by being applied to an array of tasks. These include code generation [4, 11, 13, 54], program repair [14, 26, 43] and providing explanation for code or theories [34, 36]. These broad SE application stems from their robust training on extensive code and text data, which enhances their capabilities in both linguistic understanding and code comprehension.

### 5.1 LLMs for Code Generation

Existing studies have extensively examined the capabilities of Large Language Models (LLMs) in code generation [4, 11, 13, 32, 52, 54, 58] and answer explanation [27, 28, 34]. However, these works primarily focus on evaluating correctness using automated metrics such as raw accuracy, execution rate [4], or popular evaluation metrics like pass@k [13, 52], often relying on autograders or test cases [11].

In contrast, our study goes beyond correctness by incorporating manual grading to assess the quality of LLM-generated code with respect to correctness, algorithm design, and readability. This approach allows us to evaluate whether LLMs can serve as reliable resources for assisting student while learning a new programming language and taking a intermediate-level course. Our objective is similar to the work by Yetistiren, Ozsoy, and Tuzun [58] where the authors assess the quality of Copilot-generated code by evaluating code efficiency and code validity, providing a more comprehensive analysis than many existing studies.

In assessing LLMs, our λCODEGEN benchmark involves multi-task function generation. This differs from benchmarks like HumanEval [9], MBPP [3], and HumanEval+ [31], which use short fragments (e.g., HumanEval averages 11.5 lines and 24.4 tokens per solution). While LLMs excel on these tasks, they do not capture realistic programming challenges. Some studies address this by evaluating larger code generation tasks [13], competition-level code [22], or GitHub repository code [22, 50, 52], finding lower performance on more complex problems. However, these evaluations mainly target high-resource languages like Python; our study investigates whether LLM success extends to functional programming in an educational context.

## 5.2 LLMs for Code Repair

In addition to code generation, LLMs have been widely applied to fixing buggy programs. Prior work has explored various generative program repair paradigms, including zero-shot approaches that leverage incorrect code either with accompanying instructions [14, 26, 43] or without explicit guidance [16]. In a zero-shot setting, the model is asked to perform the task without having seen any examples of correct input-output pairs beforehand. This approach is intended to reflect how a student might use an LLM to debug code without relying on pre-crafted examples or demonstrations.

The repair of semantic or logical errors in syntactically valid programs has received particular attention, with studies demonstrating performance improvements through fine-tuning [53, 59] or few-shot learning [38, 51] . Recent work has predominantly focused on ChatGPT/GPT-4o [48, 49, 56]. However, these studies focus on mainstream languages, while our study aims to understand the effectiveness for low-resource programming languages. For syntax and type errors, specialized approaches have emerged: [1] developed SYNSHINE, a RoBERTa-based system fine-tuned with compiler diagnostics, while RING [23] leverages Codex with few-shot learning for multilingual syntax repair. However, a performance gap persists between these LLM-based tools and traditional program repair systems.

Fixing type error has seen notable advances, particularly for Python. [10] introduced PyTy, a T5-based model fine-tuned on 2,766 error-fix pairs, contrasting with TypeFix's [40] zero-shot prompt-based approach. For OCaml specifically, Ribeiro, Macedo, Tsushima, Abreu, and Saraiva [44] proposed Mentat, which combines static analysis with GPT-3's generative capabilities. While these tools demonstrate progress, our work provides the first comprehensive comparison of modern LLMs across all three error categories (logical, syntactic, and type) for OCaml, revealing unique challenges in functional language repair.

### 5.3 LLMs for Conceptual Explanation

Prior work has examined LLMs for explanation in various educational contexts. AI tools like FLIP and Edison [34, 36]and other work [33] integrated LLM-generated into classroom, which students generally found helpful. And [37] uses CodeLlama to grade student submissions based on a similar grading schema to ours. Building on this foundation, recent work has established frameworks for evaluating LLM explanation quality in computer science. as shown by Kabir, Udo-Imeh, Kou, and Zhang [24] in their analysis of 517 programming questions from Stack Overflow. Their study found that while ChatGPT answers were preferred 35 % of the time for their comprehensiveness and articulate style, a concerning 52 % contained incorrect information. Ali, Rao, Mai, and Xie [2] developed an automated benchmarking approach using on the SCS1 and BDSI concept inventories on 10 LLMs. Their evaluation results demonstrate that while LLMs perform well on basic coding concepts, they struggle with more complex topics like nested conditionals, and longer questions, and runtime analysis. These limitation align with our findings where LLMs also achieve higher Mastery rate on $\lambda$Explain benchmark.

The quality of LLM explanations has been further examined through comparative studies. Lekshmi-Narayanan, Oli, Chapagain, Hassany, Banjade, Brusilovsky, and Rus [29] found that while ChatGPT's explanations showed reasonable semantic similarity to expert solutions, their lower readability posed challenges for novice learners - a pattern that aligns with the findings in [24] where 77 % of ChatGPT's answers were overly verbose. We also observed a similar trend in our evaluation of the $\lambda$Explain benchmark where explanations often suffered from excessive verbosity.

### 6  Conclusion

This work provides a comprehensive survey of the capabilities of LLMs in the context of an intermediate functional programming course. For this evaluation, we built three benchmarks $\lambda$CodeGen, $\lambda$Repair,and $\lambda$Explain and conducted a comprehensive evaluation on code generation, code repair, and conceptual explanations on 9 state-of-the-art LLMs in a functional programming course setting. Our results show that our top three LLMs, are highly effective on all tasks. Yet, LLMs are far from perfect and the best LLMs are graded as a B+/B overall. Our study serves as a snapshot of current LLMs in functional programming and a framework for ongoing evaluations as these AI tools continue to evolve. It provides insights for different audiences and stimulates various lines of future work.

### References

[1]  Toufique Ahmed, Noah Rose Ledesma, and Premkumar Devanbu. "SynShine: Improved Fixing of Syntax Errors". In: *IEEE Trans. Softw. Eng.* 49.4 (Apr. 2023), pages 2169–2181. ISSN: 0098-5589. DOI: 10.1109/TSE.2022.3212635.

[2]  Murtaza Ali, Prerna Rao, Yifan Mai, and Benjamin Xie. "Using Benchmarking Infrastructure to Evaluate LLM Performance on CS Concept Inventories: Challenges, Opportunities, and Critiques". In: *Proceedings of the 2024 ACM Conference on International Computing Education Research - Volume 1*. ICER '24. Melbourne, VIC, Australia: Association for Computing Machinery, 2024, pages 452–468. ISBN: 979-8-4007-0475-8. DOI: 10.1145/3632620.3671097.

[3]  Jacob Austin, Augustus Odena, Maxwell Nye, Maarten Bosma, Henryk Michalewski, David Dohan, Ellen Jiang, Carrie Cai, Michael Terry, Quoc Le, and Charles Sutton. *Program Synthesis with Large Language Models*. 2021. arXiv: 2108.07732 [cs.PL]. URL: https://arxiv.org/abs/2108.07732 (visited on 2025-02-10).

[4]  Alessio Buscemi. *A Comparative Study of Code Generation using ChatGPT 3.5 across 10 Programming Languages*. 2023. arXiv: 2308.04477 [cs.SE]. URL: https://arxiv.org/abs/2308.04477 (visited on 2025-02-10).

[5]  Benjamin Canou, Grégoire Henry, Çağdaş Bozman, and Fabrice Le Fessant. "Learn OCaml, An Online Learning Center for OCaml". In: *OCaml Users and Developers Workshop 2016*. Nara, Japan, Sept. 2016. URL: https://inria.hal.science/hal-01352015 (visited on 2025-02-10).

[6]  Federico Cassano, John Gouwar, Francesca Lucchetti, Claire Schlesinger, Anders Freeman, Carolyn Jane Anderson, Molly Q Feldman, Michael Greenberg, Abhinav Jangda, and Arjun Guha. "Knowledge Transfer from High-Resource to Low-Resource Programming Languages for Code LLMs". In: *Proc. ACM Program. Lang.* 8.OOPSLA2 (Oct. 2024). DOI: 10.1145/3689735.

[7]  Federico Cassano, John Gouwar, Daniel Nguyen, Sydney Nguyen, Luna Phipps-Costin, Donald Pinckney, Ming-Ho Yee, Yangtian Zi, Carolyn Jane Anderson, Molly Q Feldman, Arjun Guha, Michael Greenberg, and Abhinav Jangda. *MultiPL-E: A Scalable and Extensible Approach to Benchmarking Neural Code Generation*. 2022. arXiv: 2208.08227 [cs.LG]. URL: https://arxiv.org/abs/2208.08227 (visited on 2025-02-10).

[8]  Alana Ceci, Hanneli C. A. Tavante, Brigitte Pientka, and Xujie Si. "Data Collection for the Learn-OCaml Programming Platform: Modelling How Students Develop Typed Functional Programs". In: *Proceedings of the 52nd ACM Technical Symposium on Computer Science Education*. SIGCSE '21. Virtual Event, USA: Association for Computing Machinery, 2021, page 1341. ISBN: 978-1-4503-8062-1. DOI: 10.1145/3408877.3439579.

[9]  Mark Chen, Jerry Tworek, Heewoo Jun, Qiming Yuan, Henrique Ponde de Oliveira Pinto, Jared Kaplan, Harri Edwards, Yuri Burda, Nicholas Joseph, Greg Brockman, Alex Ray, Raul Puri, Gretchen Krueger, Michael Petrov, Heidy Khlaaf, Girish Sastry, Pamela Mishkin, Brooke Chan, Scott Gray, Nick Ryder, Mikhail Pavlov, Alethea Power, Lukasz Kaiser, Mohammad Bavarian, Clemens Winter, Philippe Tillet, Felipe Petroski Such, Dave Cummings, Matthias Plappert, Fotios Chantzis, Elizabeth Barnes, Ariel Herbert-Voss, William Hebgen Guss, Alex Nichol, Alex Paino, Nikolas Tezak, Jie Tang, Igor Babuschkin, Suchir Balaji, Shantanu Jain, William Saunders, Christopher Hesse, Andrew N. Carr, Jan Leike,

Josh Achiam, Vedant Misra, Evan Morikawa, Alec Radford, Matthew Knight, Miles Brundage, Mira Murati, Katie Mayer, Peter Welinder, Bob McGrew, Dario Amodei, Sam McCandlish, Ilya Sutskever, and Wojciech Zaremba. *Evaluating Large Language Models Trained on Code*. 2021. arXiv: 2107.03374 [cs.LG]. URL: https://arxiv.org/abs/2107.03374 (visited on 2025-12-01).

[10] Yiu Wai Chow, Luca Di Grazia, and Michael Pradel. "PyTy: Repairing Static Type Errors in Python". In: *Proceedings of the IEEE/ACM 46th International Conference on Software Engineering*. ICSE '24. Lisbon, Portugal: Association for Computing Machinery, 2024. ISBN: 979-8-4007-0217-4. DOI: 10.1145/3597503.3639184.

[11] Tristan Coignion, Clément Quinton, and Romain Rouvoy. "A Performance Study of LLM-Generated Code on Leetcode". In: *Proceedings of the 28th International Conference on Evaluation and Assessment in Software Engineering*. EASE '24. Salerno, Italy: Association for Computing Machinery, 2024, pages 79–89. ISBN: 979-8-4007-1701-7. DOI: 10.1145/3661167.3661221.

[12] Paul Denny, Viraj Kumar, and Nasser Giacaman. "Conversing with Copilot: Exploring Prompt Engineering for Solving CS1 Problems Using Natural Language". In: *Proceedings of the 54th ACM Technical Symposium on Computer Science Education V. 1*. SIGCSE 2023. Toronto ON, Canada: Association for Computing Machinery, 2023, pages 1136–1142. ISBN: 978-1-4503-9431-4. DOI: 10.1145/3545945.3569823.

[13] Xueying Du, Mingwei Liu, Kaixin Wang, Hanlin Wang, Junwei Liu, Yixuan Chen, Jiayi Feng, Chaofeng Sha, Xin Peng, and Yiling Lou. "Evaluating Large Language Models in Class-Level Code Generation". In: *Proceedings of the IEEE/ACM 46th International Conference on Software Engineering*. ICSE '24. Lisbon, Portugal: Association for Computing Machinery, 2024. ISBN: 979-8-4007-0217-4. DOI: 10.1145/3597503.3639219.

[14] Zhiyu Fan, Xiang Gao, Martin Mirchev, Abhik Roychoudhury, and Shin Hwei Tan. "Automated Repair of Programs from Large Language Models". In: *Proceedings of the 45th International Conference on Software Engineering*. ICSE '23. Melbourne, Victoria, Australia: IEEE Press, 2023, pages 1469–1481. ISBN: 978-1-6654-5701-9. DOI: 10.1109/ICSE48619.2023.00128.

[15] Laura Farinetti and Luca Cagliero. "A Critical Approach to ChatGPT: An Experience in SQL Learning". In: *Proceedings of the 56th ACM Technical Symposium on Computer Science Education V. 1*. SIGCSETS 2025. Pittsburgh, PA, USA: Association for Computing Machinery, 2025, pages 318–324. ISBN: 979-8-4007-0531-1. DOI: 10.1145/3641554.3701932.

[16] Michael Fu, Chakkrit Tantithamthavorn, Trung Le, Van Nguyen, and Dinh Phung. "VulRepair: A T5-based Automated Software Vulnerability Repair". In: *Proceedings of the 30th ACM Joint European Software Engineering Conference and Symposium on the Foundations of Software Engineering*. ESEC/FSE 2022. Singapore, Singapore: Association for Computing Machinery, 2022, pages 935–947. ISBN: 978-1-4503-9413-0. DOI: 10.1145/3540250.3549098.

[17]    Jörg von Garrel and Jana Mayer. "Artificial Intelligence in Studies—Use of ChatGPT and AI-based Tools Among Students in Germany". In: *Humanities and Social Sciences Communications* 10.1 (Nov. 2023), page 799. ISSN: 2662-9992. DOI: 10.1057/s41599-023-02304-7.

[18]    Sumit Gulwani, Ivan Radiček, and Florian Zuleger. "Automated Clustering and Program Repair for Introductory Programming Assignments". In: *Proceedings of the 39th ACM SIGPLAN Conference on Programming Language Design and Implementation*. PLDI 2018. Philadelphia, PA, USA: Association for Computing Machinery, 2018, pages 465–480. ISBN: 978-1-4503-5698-5. DOI: 10.1145/3192366.3192387.

[19]    Aliya Hameer and Brigitte Pientka. "Teaching the art of functional programming using automated grading (experience report)". In: *Proc. ACM Program. Lang.* 3.ICFP (July 2019). DOI: 10.1145/3341719.

[20]    Arto Hellas, Juho Leinonen, and Leo Leppänen. "Experiences from Integrating Large Language Model Chatbots into the Classroom". In: *Proceedings of the 2024 on ACM Virtual Global Computing Education Conference V. 1*. SIGCSE Virtual 2024. Virtual Event, NC, USA: Association for Computing Machinery, 2024, pages 46–52. ISBN: 979-8-4007-0598-4. DOI: 10.1145/3649165.3690101.

[21]    Yang Hu, Umair Z. Ahmed, Sergey Mechtaev, Ben Leong, and Abhik Roychoudhury. "Re-factoring Based Program Repair Applied to Programming Assignments". In: *Proceedings of the 34th IEEE/ACM International Conference on Automated Software Engineering*. ASE '19. San Diego, California: IEEE Press, 2020, pages 388–398. ISBN: 978-1-7281-2508-4. DOI: 10.1109/ASE.2019.00044.

[22]    Carlos E. Jimenez, John Yang, Alexander Wettig, Shunyu Yao, Kexin Pei, Ofir Press, and Karthik Narasimhan. *SWE-bench: Can Language Models Resolve Real-World GitHub Issues?* 2024. arXiv: 2310.06770 [cs.CL]. URL: https://arxiv.org/abs/2310.06770 (visited on 2025-07-10).

[23]    Harshit Joshi, José Cambronero Sanchez, Sumit Gulwani, Vu Le, Ivan Radiček, and Gust Verbruggen. "Repair Is Nearly Generation: Multilingual Program Repair with LLMs". In: *Proceedings of the Thirty-Seventh AAAI Conference on Artificial Intelligence and Thirty-Fifth Conference on Innovative Applications of Artificial Intelligence and Thirteenth Symposium on Educational Advances in Artificial Intelligence*. AAAI'23/IAAI'23/EAAI'23. AAAI Press, 2023. ISBN: 978-1-5773-5880-0. DOI: 10.1609/aaai.v37i4.25642.

[24]    Samia Kabir, David N. Udo-Imeh, Bonan Kou, and Tianyi Zhang. "Is Stack Overflow Obsolete? An Empirical Study of the Characteristics of ChatGPT Answers to Stack Overflow Questions". In: *Proceedings of the 2024 CHI Conference on Human Factors in Computing Systems*. CHI '24. Honolulu, HI, USA: Association for Computing Machinery, 2024. ISBN: 979-8-4007-0330-0. DOI: 10.1145/3613904.3642596.

[25] Hieke Keuning, Isaac Alpizar-Chacon, Ioanna Lykourentzou, Lauren Beehler, Christian Köppe, Imke de Jong, and Sergey Sosnovsky. "Students' Perceptions and Use of Generative AI Tools for Programming Across Different Computing Courses". In: *Proceedings of the 24th Koli Calling International Conference on Computing Education Research*. Koli Calling '24. New York, NY, USA: Association for Computing Machinery, 2024. ISBN: 979-8-4007-1038-4. DOI: 10.1145/3699538.3699546.

[26] Sophia D. Kolak, Ruben Martins, Claire Le Goues, and Vincent Josua Hellendoorn. "Patch Generation with Language Models: Feasibility and Scaling Behavior". In: *Deep Learning for Code Workshop*. 2022. URL: https://par.nsf.gov/biblio/10340618 (visited on 2025-02-10).

[27] Kaivaram Sivarama Krishna, Bommala Deepak Kumar, Morumpalli Deekshith Reddy, Bheema Sai Varun, and Meena Belwal. "Learning To Code With Text-Bison-001:A Beginner-Friendly Explainer for Python, C, Java". In: *2024 15th International Conference on Computing Communication and Networking Technologies (ICCCNT)*. 2024, pages 1–6. DOI: 10.1109/ICCCNT61001.2024.10724566.

[28] Juho Leinonen, Paul Denny, Stephen MacNeil, Sami Sarsa, Seth Bernstein, Joanne Kim, Andrew Tran, and Arto Hellas. "Comparing Code Explanations Created by Students and Large Language Models". In: *Proceedings of the 2023 Conference on Innovation and Technology in Computer Science Education V. 1*. ITiCSE 2023. Turku, Finland: Association for Computing Machinery, 2023, pages 124–130. ISBN: 979-8-4007-0138-2. DOI: 10.1145/3587102.3588785.

[29] Arun-Balajiee Lekshmi-Narayanan, Priti Oli, Jeevan Chapagain, Mohammad Hassany, Rabin Banjade, Peter Brusilovsky, and Vasile Rus. "Explaining Code Examples in Introductory Programming Courses: LLM vs Humans". In: *Proceedings of the 2024 AAAI Conference on Artificial Intelligence*. Edited by Muktha Ananda, Debshila Basu Malick, Jill Burstein, Lydia T. Liu, Zitao Liu, James Sharpnack, Zichao Wang, and Serena Wang. Volume 257. Proceedings of Machine Learning Research. PMLR, 26–27 Feb 2024, pages 107–117. URL: https://proceedings.mlr.press/v257/lekshmi-narayanan24a.html (visited on 2025-07-10).

[30] Fengjie Li, Jiajun Jiang, Jiajun Sun, and Hongyu Zhang. "Hybrid Automated Program Repair by Combining Large Language Models and Program Analysis". In: *ACM Trans. Softw. Eng. Methodol.* 34.7 (Aug. 2025). ISSN: 1049-331X. DOI: 10.1145/3715004.

[31] Jiawei Liu, Chunqiu Steven Xia, Yuyao Wang, and Lingming Zhang. "Is Your Code Generated by ChatGPT Really Correct? Rigorous Evaluation of Large Language Models for Code Generation". In: *Proceedings of the 37th International Conference on Neural Information Processing Systems*. NIPS '23. New Orleans, LA, USA: Curran Associates Inc., 2023. URL: https://proceedings.neurips.cc/paper_files/paper/2023/file/43e9d647ccd3e4b7b5baab53f0368686-Paper-Conference.pdf (visited on 2025-07-10).

[32] Shuai Lu, Nan Duan, Hojae Han, Daya Guo, Seung-won Hwang, and Alexey Svyatkovskiy. "ReACC: A Retrieval-Augmented Code Completion Framework". In: *Proceedings of the 60th Annual Meeting of the Association for Computational Linguistics (Volume 1: Long Papers), ACL 2022, Dublin, Ireland, May 22-27, 2022*. Edited by Smaranda Muresan, Preslav Nakov, and Aline Villavicencio. Association for Computational Linguistics, 2022, pages 6227–6240. DOI: 10.18653/V1/2022.ACL-LONG.431.

[33] Wenhan Lyu, Yimeng Wang, Tingting (Rachel) Chung, Yifan Sun, and Yixuan Zhang. "Evaluating the Effectiveness of LLMs in Introductory Computer Science Education: A Semester-Long Field Study". In: *Proceedings of the Eleventh ACM Conference on Learning @ Scale*. L@S '24. Atlanta, GA, USA: Association for Computing Machinery, 2024, pages 63–74. ISBN: 979-8-4007-0633-2. DOI: 10.1145/3657604.3662036.

[34] Stephen MacNeil, Andrew Tran, Arto Hellas, Joanne Kim, Sami Sarsa, Paul Denny, Seth Bernstein, and Juho Leinonen. "Experiences from Using Code Explanations Generated by Large Language Models in a Web Software Development E-Book". In: *Proceedings of the 54th ACM Technical Symposium on Computer Science Education V. 1*. SIGCSE 2023. Toronto, ON, Canada: Association for Computing Machinery, 2023, pages 931–937. ISBN: 978-1-4503-9431-4. DOI: 10.1145/3545945.3569785.

[35] Anders Miltner, Adrian Trejo Nuñez, Ana Brendel, Swarat Chaudhuri, and Isil Dillig. "Bottom-up synthesis of recursive functional programs using angelic execution". In: *Proc. ACM Program. Lang.* 6.POPL (Jan. 2022). DOI: 10.1145/3498682.

[36] Mihran Miroyan, Chancharik Mitra, Rishi Jain, Gireeja Ranade, and Narges Norouzi. "Analyzing Pedagogical Quality and Efficiency of LLM Responses with TA Feedback to Live Student Questions". In: *Proceedings of the 56th ACM Technical Symposium on Computer Science Education V. 1*. SIGCSETS 2025. Pittsburgh, PA, USA: Association for Computing Machinery, 2025, pages 770–776. ISBN: 979-8-4007-0531-1. DOI: 10.1145/3641554.3701965.

[37] Goda Nagakalyani, Saurav Chaudhary, Varsha Apte, Ganesh Ramakrishnan, and Srikanth Tamilselvam. "Design and Evaluation of an AI-Assisted Grading Tool for Introductory Programming Assignments: An Experience Report". In: *Proceedings of the 56th ACM Technical Symposium on Computer Science Education V. 1*. SIGCSETS 2025. Pittsburgh, PA, USA: Association for Computing Machinery, 2025, pages 805–811. ISBN: 979-8-4007-0531-1. DOI: 10.1145/3641554.3701913.

[38] Noor Nashid, Mifta Sintaha, and Ali Mesbah. "Retrieval-Based Prompt Selection for Code-Related Few-Shot Learning". In: *Proceedings of the 45th International Conference on Software Engineering*. ICSE '23. Melbourne, Victoria, Australia: IEEE Press, 2023, pages 2450–2462. ISBN: 978-1-6654-5701-9. DOI: 10.1109/ICSE48619.2023.00205.

[39]   Peter-Michael Osera and Steve Zdancewic. "Type-and-Example-Directed Program Synthesis". In: PLDI '15. Portland, OR, USA: Association for Computing Machinery, 2015, pages 619–630. ISBN: 978-1-4503-3468-6. DOI: 10.1145/2737924.2738007.

[40]   Yun Peng, Shuzheng Gao, Cuiyun Gao, Yintong Huo, and Michael Lyu. "Domain Knowledge Matters: Improving Prompts with Fix Templates for Repairing Python Type Errors". In: *Proceedings of the IEEE/ACM 46th International Conference on Software Engineering*. ICSE '24. Lisbon, Portugal: Association for Computing Machinery, 2024. ISBN: 979-8-4007-0217-4. DOI: 10.1145/3597503.3608132.

[41]   Siddhartha Prasad, Ben Greenman, Tim Nelson, and Shriram Krishnamurthi. "Generating Programs Trivially: Student Use of Large Language Models". In: *Proceedings of the ACM Conference on Global Computing Education Vol 1*. CompEd 2023. Hyderabad, India: Association for Computing Machinery, 2023, pages 126–132. ISBN: 979-8-4007-0048-4. DOI: 10.1145/3576882.3617921.

[42]   James Prather, Juho Leinonen, Natalie Kiesler, Jamie Gorson Benario, Sam Lau, Stephen MacNeil, Narges Norouzi, Simone Opel, Vee Pettit, Leo Porter, Brent N. Reeves, Jaromir Savelka, David H. Smith, Sven Strickroth, and Daniel Zingaro. "Beyond the Hype: A Comprehensive Review of Current Trends in Generative AI Research, Teaching Practices, and Tools". In: *2024 Working Group Reports on Innovation and Technology in Computer Science Education*. ITiCSE 2024. Milan, Italy: Association for Computing Machinery, 2025, pages 300–338. ISBN: 979-8-4007-1208-1. DOI: 10.1145/3689187.3709614.

[43]   Julian Aron Prenner, Hlib Babii, and Romain Robbes. "Can OpenAI's Codex Fix Bugs? An Evaluation on QuixBugs". In: *Proceedings of the Third International Workshop on Automated Program Repair*. APR '22. Pittsburgh, Pennsylvania: Association for Computing Machinery, 2022, pages 69–75. ISBN: 978-1-4503-9285-3. DOI: 10.1145/3524459.3527351.

[44]   Francisco Ribeiro, José Nuno Castro de Macedo, Kanae Tsushima, Rui Abreu, and João Saraiva. "GPT-3-Powered Type Error Debugging: Investigating the Use of Large Language Models for Code Repair". In: *Proceedings of the 16th ACM SIGPLAN International Conference on Software Language Engineering*. SLE 2023. Cascais, Portugal: Association for Computing Machinery, 2023, pages 111–124. ISBN: 979-8-4007-0396-6. DOI: 10.1145/3623476.3623522.

[45]   Georgios Sakkas, Madeline Endres, Benjamin Cosman, Westley Weimer, and Ranjit Jhala. "Type Error Feedback via Analytic Program Repair". In: *Proceedings of the 41st ACM SIGPLAN Conference on Programming Language Design and Implementation*. PLDI 2020. London, UK: Association for Computing Machinery, 2020, pages 16–30. ISBN: 978-1-4503-7613-6. DOI: 10.1145/3385412.3386005.

[46]   Sofia Santos, João Saraiva, and Francisco Ribeiro. "Large Language Models in Automated Repair of Haskell Type Errors". In: *Proceedings of the 5th ACM/IEEE International Workshop on Automated Program Repair*. APR '24. Lisbon, Portugal: Association for Computing Machinery, 2024, pages 42–45. ISBN: 979-8-4007-0577-9. DOI: 10.1145/3643788.3648012.

[47] Sami Sarsa, Paul Denny, Arto Hellas, and Juho Leinonen. "Automatic Generation of Programming Exercises and Code Explanations Using Large Language Models". In: *Proceedings of the 2022 ACM Conference on International Computing Education Research - Volume 1*. ICER '22. Lugano and Virtual Event, Switzerland: Association for Computing Machinery, 2022, pages 27–43. ISBN: 978-1-4503-9194-8. DOI: 10.1145/3501385.3543957.

[48] Dominik Sobania, Martin Briesch, Carol Hanna, and Justyna Petke. "An Analysis of the Automatic Bug Fixing Performance of ChatGPT". In: *2023 IEEE/ACM International Workshop on Automated Program Repair (APR)*. Los Alamitos, CA, USA: IEEE Computer Society, May 2023, pages 23–30. DOI: 10.1109/APR59189.2023.00012.

[49] Nigar Surameery and Mohammed Shakor. "Use Chat GPT to Solve Programming Bugs". In: *International Journal of Information Technology and Computer Engineering* (Jan. 2023), pages 17–22. DOI: 10.55529/ijitc.31.17.22.

[50] Priyan Vaithilingam, Tianyi Zhang, and Elena L. Glassman. "Expectation vs. Experience: Evaluating the Usability of Code Generation Tools Powered by Large Language Models". In: *Extended Abstracts of the 2022 CHI Conference on Human Factors in Computing Systems*. CHI EA '22. New Orleans, LA, USA: Association for Computing Machinery, 2022. ISBN: 978-1-4503-9156-6. DOI: 10.1145/3491101.3519665.

[51] Tung Do Viet and Konstantin Markov. "Using Large Language Models for Bug Localization and Fixing". In: *2023 12th International Conference on Awareness Science and Technology (iCAST)*. 2023, pages 192–197. DOI: 10.1109/iCAST57874.2023.10359304.

[52] Chong Wang, Jian Zhang, Yebo Feng, Tianlin Li, Weisong Sun, Yang Liu, and Xin Peng. "Teaching Code LLMs to Use Autocompletion Tools in Repository-Level Code Generation". In: *ACM Trans. Softw. Eng. Methodol.* 34.7 (Aug. 2025). ISSN: 1049-331X. DOI: 10.1145/3714462.

[53] Weishi Wang, Yue Wang, Shafiq Joty, and Steven C.H. Hoi. "RAP-Gen: Retrieval-Augmented Patch Generation with CodeT5 for Automatic Program Repair". In: *Proceedings of the 31st ACM Joint European Software Engineering Conference and Symposium on the Foundations of Software Engineering*. ESEC/FSE 2023. San Francisco, CA, USA: Association for Computing Machinery, 2023, pages 146–158. ISBN: 979-8-4007-0327-0. DOI: 10.1145/3611643.3616256.

[54] Michel Wermelinger. "Using GitHub Copilot to Solve Simple Programming Problems". In: *Proceedings of the 54th ACM Technical Symposium on Computer Science Education V. 1*. SIGCSE 2023. Toronto ON, Canada: Association for Computing Machinery, 2023, pages 172–178. ISBN: 978-1-4503-9431-4. DOI: 10.1145/3545945.3569830.

[55] An Yang, Baosong Yang, Beichen Zhang, Binyuan Hui, Bo Zheng, Bowen Yu, Chengyuan Li, Dayiheng Liu, Fei Huang, Haoran Wei, Huan Lin, Jian Yang, Jianhong Tu, Jianwei Zhang, Jianxin Yang, Jiaxi Yang, Jingren Zhou, Junyang Lin, Kai Dang, Keming Lu, Keqin Bao, Kexin Yang, Le Yu, Mei Li, Mingfeng Xue,

Pei Zhang, Qin Zhu, Rui Men, Runji Lin, Tianhao Li, Tianyi Tang, Tingyu Xia, Xingzhang Ren, Xuancheng Ren, Yang Fan, Yang Su, Yichang Zhang, Yu Wan, Yuqiong Liu, Zeyu Cui, Zhenru Zhang, and Zihan Qiu. *Qwen2.5 Technical Report*. 2025. arXiv: 2412.15115 [cs.CL]. URL: https://arxiv.org/abs/2412.15115 (visited on 2025-02-10).

[56]  Boyang Yang, Haoye Tian, Weiguo Pian, Haoran Yu, Haitao Wang, Jacques Klein, Tegawendé F. Bissyandé, and Shunfu Jin. "CREF: An LLM-Based Conversational Software Repair Framework for Programming Tutors". In: *Proceedings of the 33rd ACM SIGSOFT International Symposium on Software Testing and Analysis*. ISSTA 2024. Vienna, Austria: Association for Computing Machinery, 2024, pages 882–894. ISBN: 979-8-4007-0612-7. DOI: 10.1145/3650212.3680328.

[57]  Stephanie Yang, Hanzhang Zhao, Yudian Xu, Karen Brennan, and Bertrand Schneider. "Debugging with an AI Tutor: Investigating Novice Help-seeking Behaviors and Perceived Learning". In: *Proceedings of the 2024 ACM Conference on International Computing Education Research - Volume 1*. ICER '24. Melbourne, VIC, Australia: Association for Computing Machinery, 2024, pages 84–94. ISBN: 979-8-4007-0475-8. DOI: 10.1145/3632620.3671092.

[58]  Burak Yetistiren, Isik Ozsoy, and Eray Tuzun. "Assessing the Quality of GitHub Copilot's Code Generation". In: *Proceedings of the 18th International Conference on Predictive Models and Data Analytics in Software Engineering*. PROMISE 2022. Singapore, Singapore: Association for Computing Machinery, 2022, pages 62–71. ISBN: 978-1-4503-9860-2. DOI: 10.1145/3558489.3559072.

[59]  Wei Yuan, Quanjun Zhang, Tieke He, Chunrong Fang, Nguyen Quoc Viet Hung, Xiaodong Hao, and Hongzhi Yin. "CIRCLE: Continual Repair across Programming Languages". In: *Proceedings of the 31st ACM SIGSOFT International Symposium on Software Testing and Analysis*. ISSTA 2022. Virtual, South Korea: Association for Computing Machinery, 2022, pages 678–690. ISBN: 978-1-4503-9379-9. DOI: 10.1145/3533767.3534219.

[60]  Yihan Zhang, Brigitte Pientka, and Xujie Si. *Artifact for: Evaluating LLMs in the Context of a Functional Programming Course: A Comprehensive Study*. Zenodo. 2026. DOI: 10.5281/zenodo.18470483.

[61]  Qinkai Zheng, Xiao Xia, Xu Zou, Yuxiao Dong, Shan Wang, Yufei Xue, Lei Shen, Zihan Wang, Andi Wang, Yang Li, Teng Su, Zhilin Yang, and Jie Tang. "CodeGeeX: A Pre-Trained Model for Code Generation with Multilingual Benchmarking on HumanEval-X". In: *Proceedings of the 29th ACM SIGKDD Conference on Knowledge Discovery and Data Mining*. KDD '23. Long Beach, CA, USA: Association for Computing Machinery, 2023, pages 5673–5684. ISBN: 979-8-4007-0103-0. DOI: 10.1145/3580305.3599790.

## About the authors

**Yihan Zhang** is a Master student in McGill University, Canada. Contact her at yihan.zhang2@mail.mcgill.ca.
https://orcid.org/0009-0002-9107-6774

**Brigitte Pientka** is a professor in McGill University, Canada. Contact her at brigitte.pientka@mcgill.ca.
https://orcid.org/0000-0002-2549-4276

**Xujie Si** is a professor in University of Toronto, Canada. Contact him at six@cs.toronto.edu.
https://orcid.org/0000-0002-3739-2269